\documentclass[12pt,a4paper,twoside]{article}

\newcommand{\be}{\begin{equation}}
\newcommand{\ee}{\end{equation}}
\newcommand{\beqs}{\begin{eqnarray}}
\newcommand{\eeqs}{\end{eqnarray}}

\newcommand{\half}{{1 \over 2}}

\expandafter\ifx\csname mathbbm\endcsname\relax

\else

\fi
\begin{document}
\begin{titlepage}
\begin{flushleft}  
       \hfill                      {\tt hep-th/0304135}\\
       \hfill                       April 2003\\
\end{flushleft}
\vspace*{3mm}
\begin{center}
{\Large  Area spectrum and quasinormal modes of black holes\\}
\vspace*{12mm}
\large Alexios P. Polychronakos\footnote{On leave from Physics Department,
University of Ioannina, Greece.} \\
\vspace*{5mm}
{\em Physics Department, Baruch College of CUNY\\
New York, NY 10010, USA\\
\small alexios@gursey.baruch.cuny.edu \/}\\
\vspace*{4mm}
\vspace*{15mm}
\end{center}

\begin{abstract}
We demonstrate that an equidistant area spectrum for the link variables
in loop quantum gravity can reproduce both the thermodynamics and 
the quasinormal mode properties of black holes.

\end{abstract}

\end{titlepage}

\section{Introduction}

The quantum properties of black holes is a fascinating subject. Starting
with the pioneering work of Bekenstein and Hawking \cite{Bek,Haw}, 
and continuing into the realms of string theory and quantum gravity
\cite{Sus,ABCK}, they have been the subject of intense investigation.

An important issue is the extent to which the fundamental quantum and
statistical mechanical properties of black holes are fixed by internal
consistency and their known macroscopic properties alone; that is, 
the extent to which these properties are independent of any specific
model or theory of quantum gravity.

Interesting steps in this direction were taken recently by semiclassical
consideration of the macroscopic oscillation modes of black holes, in an
approach reminiscent to Debye's theory for the heat capacity of solids. 
Hod \cite{Hod} has pointed out that the decaying `ringing modes' 
(or quasinormal modes) of a Schwartzchild black
hole conform with the existence of the Bekenstein-Hawking temperature
and imply emission in terms of a fixed area quantum. He further noticed
that, in Planck units ($c=\hbar=G=1$), this quantum has the numerical value
$4 \ln 3$ (a fact later proved analytically \cite{Mot}). This suggests
a composition of the black hole in terms of three-state objects. 
Dreyer \cite{Dre} proposed
that this result could be recovered in loop quantum gravity if the relevant
group is taken to be $SO(3)$ rather than $SU(2)$. These developments
have spurred a lot of subsequent activity \cite{Gab}-\cite{CL}.

Although the above considerations are still somewhat speculative, they
certainly point to the intriguing possibility of using classical,
macroscopic properties of black holes to infer information on their
quantum, microscopic nature. It is, therefore, of interest to explore
the alternative microscopic scenaria in which these properties 
can be accomodated. This is the subject of this paper. As we shall see,
an equidistant spectrum naturally reproduces thermodynamic and ringing
mode properties of black holes, still allowing for at least two distinct
possibilities, depending on the quantum statistics of their microscopic
constituents.

\section{The quasinormal mode argument}

We summarize here the connection between the decaying ringing modes
of a black hole and its area spectrum \cite{Hod,Dre}. 

Perturbations in the metric of a black hole of mass $M$ decay according
to a set of characteristic discrete (complex) frequencies. These 
frequencies for highly damped modes are known to asymptote to the values
\cite{Nol,KS,Hod,Mot}
\be
\omega_n = \frac{\ln3}{8\pi M} -i \frac{2\pi}{8\pi M} 
\left( n+\half \right)
\ee
The spacing of the imaginary part of these frequencies agrees with
the Bekenstein-Hawking temperature of the black hole $T_{BH} = 1/(8\pi M)$. 
Hod suggested that the
asymptotically constant real part can be semiclassically interpreted
as the quantum of energy $\Delta E = \Delta M$ emitted by the black hole. 
Using the standard expression for the area of the black hole
$A = 16 \pi M^2$, this implies that the area decreases by a quantum
$\Delta A$:
\be
\Delta A = 32 \pi M \Delta M = 4 \ln 3
\ee
Hod further observed that, if we consider the black hole as composed
of $N$ `bits', each contributing a quantum of area $4 \ln 3$ and having
3 internal states, then the total number of states of the black hole
would be $\Omega = 3^N$ and the entropy would calculate to
\be
S = \ln \Omega = N \ln 3 = \frac{1}{4} (N 4\ln3 ) = \frac{1}{4} A
\ee
in agreement with the Bekenstein-Hawking area-entropy result.

Finally, Dreyer suggested that the three-state `bits' can be spin-1
link representations in the loop quantum gravity formulation. To
have the black hole dominated by spin-1 links (rather than 
spin-$\half$ ones, which is the standard result), Dreyer proposed a
modification of the link group from $SU(2)$ to $SO(3)$; then only 
integer-spin representations are allowed, with the lowest nontrivial
one, spin-1, dominating.

Discarding half-integer spins can be unwanted, since they are
required if the back hole is to couple to half-integer spin (fermionic)
fields. Further, Dreyer's analysis assumes the standard expression for 
the area and degeneracy contribution of each link to the black hole.
Alternative approaches, suggesting an equidistant link area spectrum
and a different state counting have been proposed. We shall revisit
this analysis by considering these different possibilities.

\section{The composite black hole model}

The spectrum of the area of black holes, and its degeneracy $\Omega$
for a given area, are the important ingredients in their quantum 
statistics. Their determination involves some version of a quantum
gravity model for the black hole.

In a large class of quantum gravity models,
black holes are effectively represented as composite systems consisting
of a large number of identical components. Each component contributes
an elementary area, and the total area of the black hole horizon
is the sum of the areas of the components. The full spectrum and degeneracy
of the black hole area is determined by the area spectrum and degeneracies
of each component, as well as the quantum statistics obeyed by the
(identical, but not necessarily indistinguishable) components.

As it turns out, {\it any} individual component spectrum and degeneracy will
lead to the area-entropy law $S\sim A$ for a macroscopic black hole, 
as long as the components are considered to be {\it somewhat} 
distinguishable \cite{APS}. Completely indistinguishable statistics 
for the components (fermionic or bosonic), 
on the other hand, lead to the weaker entropy dependence $S \sim A^c$, 
with the exponent $c<1$ depending on the area spectrum of the components.

In this paper we shall concentrate on the composite model deriving from
loop quantum gravity \cite{RS,AL,ABCK}. 
In it, the components are horizon-piercing links 
on a random lattice, each carrying a representation of $SU(2)$. As such, 
their elementary area depends on the quadratic Casimir of the representation, 
while the degeneracy is the dimensionality of the representation. 
Labeling the spin $j$ of the representation by the integer $n=2j=0,1,2,\dots$,
the area spectrum of each component is $a_n$ and its degeneracy is $g_n = n+1$.

Strictly speaking, the link states are not truly independent, since there
is the Gauss law constraint that the total $SU(2)$ state of the black hole
should be a singlet. The ensuing reduction of the state space, however, 
proves to be irrelevant for the thermodynamics of macroscopic black holes;
in this limit, the link components can be treated as independent.

There are different proposals for the form of the spectrum $a_n$. The
standard result \cite{RS} is that the area is proportional to the square 
root of the Casimir, that is,
\be
a_n = 2\gamma \sqrt{j(j+1)} = \gamma \sqrt{n(n+2)}
\label{sqrt}\ee
with $\gamma$ a proportionality constant (related to the so-called
Immirzi parameter). 

In \cite{APS}, on the other
hand, it was argued that quantum corrections would additively renormalize 
the Casimir into $j(j+1) + \frac{1}{4} = (j+\half)^2$, which leads to an 
{\it equidistant} spectrum:
\be
a_n = 2\gamma (j+\half) = \gamma (n+1)
\label{APSarea}\ee
This conforms with the original arguments of Bekenstein \cite{Bek}
that the horizon
area, behaving as an adiabatic invariant, should be integrally quantized,
as well as with various more recent proposals for an equidistant area 
spectrum \cite{BM}-\cite{BDK}.

There are also differing considerations of the statistics obeyed by the
components. One of them considers them as partially distinguishable, 
while another considers them as fully distinguishable. These lead to
different statistical ensembles with important physical implications.
We shall analyze the alternatives in the sequel.

\section{Partially distinguishable components}

The degeneracy of states of a black hole consisting of many components,
according to one counting \cite{ABCK}, is
\be
\Omega = \prod_{n=0}^\infty g_n^{N_n}
\label{Omegapartial}\ee
where $N_n$ is the number of components with spin $j=n/2$ and $g_n = n+1$
the number of internal states of each such component. As we shall explain
in the next section,
this assigns the components a peculiar {\it partial} distinguishability.

The statistical mechanics of a macroscopic black hole can be obtained in a 
standard way by considering a statistical ensemble of black holes, 
corresponding to an intensive `temperature' parameter $\beta$ dual to the
area (this is not the standard temperature parameter, which is dual to
the energy). The area partition function becomes
\be
Z = \sum_A \Omega (A) e^{-\beta A} = \sum_{\{ N_n \}} \prod_{n=0}^\infty 
g_n^{N_n} e^{-\beta N_n a_n}
= \prod_{n=0}^\infty \frac{1}{1-g_n e^{-\beta a_n}}
\label{Zpartial}\ee
Note that we also sum over different link numbers, the links being quantum
variables that can be created or annihilated.

For a macroscopic black hole the area and degeneracy of states grow large
and thus the above partition function must diverge. As we increase the
`temperature', that is, decrease $\beta$ from infinity, a macroscopic
black hole will form at some critical $\beta_c$ for which $Z$ diverges. 
Generically, this happens when one of the denominators in (\ref{Zpartial}),
corresponding to a specific spin $n_c$, diverges; that is,
\be
\beta_c = max\{\beta_n\} \equiv \beta_{n_c}~,~~~~ 
{\rm where}~~ \beta_n = \frac{\ln g_n}{a_n} = \frac{\ln (n+1)}{a_n}
\ee
At this point a condensate of spins $j_c = n_c /2$ forms, while the number
of other spins remains finite and of order one. The area and entropy 
are dominated by the spins of the condensate and we have
\be
A= N a_{n_c} ~,~~~ S=\ln \Omega = \ln ( g_{n_c}^N )
\ee
where $N$ is the total number of spins in the condensate. The above lead
to the entropy law
\be
S = \frac{\ln g_{n_c}}{a_{n_c}} A
\ee
Conformity with the Bekenstein-Hawking (BH) result $S= \frac{1}{4} A$ 
requires
\be
a_{n_c} = 4 \ln g_{n_c}
\ee
Transitions of this black hole to one with different area will be 
statistically dominated by processes increasing or decreasing the number of
spins in the condensate. That is, the emission spectrum of the black hole
will exhibit the area quantum
\be
\Delta A = a_{n_c}
\ee
Conformity with the ringing mode calculation $\Delta A = 4 \ln 3$, then, 
implies
\be
g_{n_c} =3 ~,~~~ a_{n_c} = 4 \ln 3
\ee
corresponding to $n=2$ (for $g_n = n+1$). This picks spin-1 links as the
ones dominating the composition of the black hole.

With the standard area spectrum (\ref{sqrt}), $a_n = \gamma \sqrt{n(n+2)}$,
the critical point is at $\beta_c = \beta_1 = \ln2 / \gamma\sqrt{3}$. 
(Note that spin-0 links, which do not contribute to the area, are
considered as part of the vacuum and omitted from (\ref{Zpartial})). This
picks spin-$\half$ components for the condensate and $g_{n_c} =2$, 
thus leading to a disagreement with the ringing mode calculation. 
The remedy proposed
for this problem is to accept only integer spins in the spectrum; that is,
consider representations of $SO(3)$ rather than $SU(2)$ \cite{Dre}. As stated
before, this has the drawback that the discarded half-integer spins
are desirable in order to have the black hole couple to fermion fields.

The first main observation of this paper is that the modified, equidistant
spectrum (\ref{APSarea}), $a_n = \gamma (n+1)$,
remedies this problem {\it without} discarding 
half-integer spins. Indeed, with this spectrum we have 
\be
\beta_0 = 0 ~,~~~ \beta_1 = \beta_3 = \frac{\ln 2}{2 \gamma} = 
\frac{0.3466}{\gamma}
~,~~~ \beta_2 = \frac{\ln 3}{3 \gamma} = \frac{0.3662}{\gamma} ~\dots
\ee
We see that the critical parameter is now shifted to $\beta_2$, which
picks spin-1 components for the condensate. Note, further, that although
spin-0 links now contribute an amount $a_1 = \gamma$ to the area and thus 
cannot be considered as part of the vacuum, they are thermodynamically 
suppressed. Thus, we obtain agreement with the ringing mode calculation 
while retaining the full $SU(2)$ spectrum.

\section{Fully distinguishable components}

The state-counting formula (\ref{Omegapartial}) corresponds to a peculiar, 
partially distinguishable statistics for the components. 
To make this point clear, consider two links with spins $j_1$ and $j_2$ 
and internal states (eigenvalues of the third component) $m_1$ and $m_2$ 
respectively. Formula (\ref{Omegapartial})
counts this as a single state, irrespective of which link carries which
spin. So the quantum state is invariant under permutation of these two
spins:
\be
| (j_1,m_1) , (j_2,m_2) \rangle = | (j_2,m_2) , (j_1,m_1) \rangle 
\label{different}\ee
On the other hand, for two equal spins $j$, formula (\ref{Omegapartial}) 
counts {\it all} $(2j+1)\times (2j+1)$ states as distinct; that is
\be
| (j,m_1) , (j,m_2) \rangle \neq | (j,m_2) , (j,m_1) \rangle 
\label{same}\ee

We see no justification for such a counting. If, indeed, nothing else
distinguishes the links except their spin, then quantum mechanics implies
that they should be treated as indistinguishable objects, in which case
(\ref{same}) is incorrect and the entropy-area law cannot be recovered.
In \cite{APS}, however, we have argued that the links are {\it fully distinct}
because of their connectivity to the lattice representing spacetime 
outside of the black hole. In this case, they should be treated as
completely distinguishable objects. (A similar point of view was put forth
by Krasnov in \cite{Kra}, and also by Rovelli \cite{Rov}, but was apparently 
abandoned in subsequent work.)

This leads to different thermodynamics. The partition function of $N$
distinguishable links is
\be
Z_N = Z_1^N ~,~~~ {\rm where} ~~ Z_1 = \sum_{n=0}^\infty g_n e^{-\beta a_n}
\ee
The full partition function for all possible numbers of links is, thus,
\be
Z = \sum_{N=0}^\infty Z_N = \frac{1}{1-Z_1}
\ee
A macroscopic black hole will, again, form when $Z$ diverges. This will
happen for the critical value $\beta_c$ for which
\be
Z_1 (\beta_c ) = 1
\ee
Standard thermodynamic arguments lead to the expression for the entropy
\be
S = \beta A + \ln Z
\label{Sdist}\ee
At the critical point, $Z$ is of order $A$; thus, $\ln Z$ is a sub-leading
logarithmic correction. Such corrections will be present, but cannot be
reliably calculated in the ensemble formulation since this implies
fluctuations for the total black hole area which would not be there in a true
area eigenstate. The leading (thermodynamic) contribution, however, is
accurately given by the first term in (\ref{Sdist}) as
\be
S = \beta_c \, A
\ee
Conformity with the standard BH result implies $\beta_c = \frac{1}{4}$.

It should be stressed that, now, the black hole state is {\it not} dominated
by any single spin value; rather, there is a Boltzmann distribution of spins 
for  all possible values. In fact, the probability that any given link 
carries spin $j=n/2$ is given by
\be
P_n = g_n e^{-\beta_c a_n} = (n+1) e^{-\beta_c a_n}
\ee

Transition to a black hole of different (lesser) area would involve transitions
between different values of spin, as well as the disappearence of links. 
Therefore, black hole transitions involve quanta of area $\Delta A =
a_n - a_{n'}$, corresponding to transitions, as well as $\Delta A = a_n$,
corresponding to annihilation of links. Note, further, that spin-0 links
{\it cannot} contribute zero area in the fully distinguishable picture;
this would lead to a thermodynamic collapse of the black hole into a state
with an arbitrarily large number of spin-0 links and arbitrarily large
entropy.

With the above conditions, it is easy to see that the standard spectrum
$a_n = \gamma \sqrt{n(n+2)}$ is inconsistent with the quantization of area 
implied by the ringing mode calculation, since it leads to an essentially 
continuous area emission spectrum with no apparent quantum. The equidistant
spectrum $a_n = \gamma (n+1)$, however, is consistent, predicting
a quantum of area $\Delta A = \gamma$. This fixes $\gamma = 4\ln 3$. 
An explicit calculation of $Z_1$, on the other hand, leads to
\be
\beta_c = \gamma^{-1} \ln \frac{3+\sqrt{5}}{2} = \frac{0.876}{4}
\ee
This differs (although not by much) from the value 
$\beta_c = \frac{1}{4}$ required by the BH area-entropy law. 

The only possibility, therefore, is that the spin content is modified.
We shall, again, only consider changing the group from $SU(2)$ to $SO(3)$, 
seeing no justification for any other modification. Thus, we now label
the states in terms of the integer spin $j=n$. The degeneracy is
$g_n = 2n+1$. For the spectrum, we must consider a renormalization of
the standard result consistent with the requirement for a quantum of
area, as explained two paragraphs above. The result (\ref{APSarea}),
$a_n = 2\gamma (j+\half) = \gamma (2n+1)$, is not satisfactory, since it 
predicts an area quantum $2\gamma$ for transitions but a different area
quantum $\gamma$ for creation or annihilation of links (for spin-0
links). We thus propose the modified equidistant spectrum
\be
a_n = \gamma (n+1) ~,~~~ g_n = 2n+1 ~;~~~ \gamma = 4\ln 3
\ee
This corresponds to a quantum $\gamma = 4\ln 3$ for transitions between 
different integer spins, and the same quantum for the creation or 
annihilation of links. 

We should stress that the above spectrum is determined only by the requirement
of a fixed area quantum. It is still not guaranteed that it will reproduce
the correct black holes thermodynamics. To that end, we
calculate the single-link partition function:
\be
Z_1 = \sum_{n=0}^\infty (2n+1) e^{-\beta \gamma (n+1)} =
\frac{e^{\beta \gamma} +1}{(e^{\beta \gamma} -1)^2}
\ee
The critical point $Z_1 =1$ is now at $e^{\beta \gamma} =3$, or
\be
\beta_c = \frac{\ln 3}{\gamma} = \frac{1}{4}
\ee
This leads to the correct BH area-entropy relation. We therefore conclude
that distinguishable components with an equidistant spectrum of integer
spins correctly reproduce both the thermodynamics and the ringing mode
properties of black holes. No single spin value dominates. The probabilities
$P_n$ of appearence of the value $j=n$ for the spin of a link are
\be
P_0 = \frac{1}{3} ~,~~~ P_1 = \frac{1}{3} ~,~~~ P_2 = \frac{5}{27} \dots
\ee

\section{Conclusions and discussion}

To summarize, we have made two main points:

1. If the standard counting formula for states (\ref{Omegapartial}) is assumed,
then the equidistant area spectrum as proposed in \cite{APS} naturally explains
the domination of spin-1 links and reproduces the ringing mode properties of
black holes, without the need to eliminate half-integer spins.

2. If totally distinguishable link statistics are assumed, then an equidistant
area spectrum of components, carrying {\it integer} spins only, correctly 
reproduces ringing mode properties, without domination of spin-1 components.

Which, if any, of the above cases is realized must be the subject of further
investigation. The successful model should account in a natural way for the
properties of black holes carrying charge and angular momentum.

A criticism for an equidistant area spectrum, as advocated in this paper,
is that it leads to substantial deviations from the black-body
spectrum of Hawking radiation. Indeed, an area and energy quantum implies
a discrete emission spectrum with the spacing of spectral lines being of
the same order of magniture as the black body thermal frequency. 

In fact, the same criticism
would apply to the standard loop quantum gravity spectrum and degeneracy
(\ref{sqrt}) and (\ref{Omegapartial}). It has been argued that the
irrational character of the area eigenvalues (\ref{sqrt}) creates a
spread of possible area quanta under transitions which effectively
reproduce a continuum. However, the state of the black hole is dominated
by a condensate of spin-$\half$ links, and the number of links with 
different spin values is of order one. All transitions of the black hole 
are dominated by processes increasing or decreasing the number of links
in the condensate and thus reproduce a discrete emission spectrum. It is
unreasonable to expect that the remaining few links account for the
bulk of the (continuous) Hawking radiation.

Distinguishable statistics actually cure that, since now the black hole
state contains all spins. This still leaves us with a disagreement
with the ringing mode result, as stated in the previous section.
It seems that to take the ringing mode connection seriously, we must
admit an equidistant area spectrum and thus accept the fact that the
Hawking radiation spectrum is discrete.

We do not feel that this is damning. The high-frequency exponential
part of the spectrum is accurately reproduced, the discreteness there
being inconsequential. This is the energy range in which the photons
(or other emitted particles)
behave essentially like classical particles, whose scattering properties
are expected to be accurately reproduced by the classical black hole
metric. For frequencies close to the thermal frequency, however, the
wavelength of the photons becomes comparable to the size of the black
hole and they sense global properties of its geometry. Back-reaction due to
geometry change at emission and absorption of such photons is expected 
to be important, the energy of these photons being of the same order
as the energy spacing of the black hole. A deviation from ideal
black-body spectrum, which assumes a fixed metric and ignores back-reaction,
would seem reasonable.

In summary, the essence of the Hawking radiation argument is that a
black hole would be in thermal equilibrium (although an unstable one)
with a heat bath at temperature $T_{BH} = 8 \pi M$. Removal of the
heat bath would deprive the black hole of all absorption, leaving only
the emitted Hawking radiation. This does not necessarily mean that
radiation is black-body. A large atom can be in thermal equilibrium
with a heat bath of some temperature, but in the abscence of the heat
bath it would radiate in its own line spectrum. Perhaps we are facing
a similar situation.

\end{document}